\documentclass[a4paper,11pt]{article}
\pdfoutput=1 

\usepackage{jinstpub} 
\usepackage{amsmath}

\usepackage{color}

\author[n]{P.~Abratenko}   
\author[g]{R.~Acciarri}
\author[bb]{C.~Adams}
\author[h]{R.~An}
\author[c]{J.~Anthony}
\author[y]{J.~Asaadi}
\author[a]{M.~Auger}
\author[g]{L.~Bagby}
\author[bb]{S.~Balasubramanian}
\author[g]{B.~Baller}
% ...

\affiliation[a]{Universit{\"a}t Bern, Bern CH-3012, Switzerland}
% ...
\affiliation[c]{University of Cambridge, Cambridge CB3 0HE, United Kingdom}
% ...
\affiliation[g]{Fermi National Accelerator Laboratory (FNAL), Batavia, IL 60510, USA}
\affiliation[h]{Illinois Institute of Technology (IIT), Chicago, IL 60616, USA}
% ...
\affiliation[n]{University of Michigan, Ann Arbor, MI, 48109, USA}
% ...
\affiliation[y]{University of Texas, Arlington, TX, 76019, USA}
% ...
\affiliation[bb]{Yale University, New Haven, CT, 06520, USA}

\title{Determination of muon momentum in the MicroBooNE LArTPC using an improved model of multiple Coulomb scattering}

\abstract{
We discuss a technique for measuring a charged particle's momentum by means of multiple Coulomb scattering (MCS) in the MicroBooNE liquid argon time projection chamber (LArTPC). This method does not require the full particle ionization track to be contained inside of the detector volume as other track momentum reconstruction methods do (range-based momentum reconstruction and calorimetric momentum reconstruction). We motivate use of this technique, describe a tuning of the underlying phenomenological formula, quantify its performance on fully contained beam-neutrino-induced muon tracks both in simulation and in data, and quantify its performance on exiting muon tracks in simulation. 
Using simulation, we have shown that the standard Highland formula should be re-tuned specifically for scattering in liquid argon, which significantly improves the bias and resolution of the momentum measurement.
With the tuned formula, we find agreement between data and simulation for contained tracks, with a small bias in the momentum reconstruction and with resolutions that vary as a function of track length, improving from about 10\% for the shortest (one meter long) tracks to 5\% for longer (several meter) tracks. For simulated exiting muons with at least one meter of track contained, we find a similarly small bias, and a resolution which is less than 15\% for muons with momentum below 2 GeV/c.  
Above 2 GeV/c, results are given as a first estimate of the MCS momentum measurement capabilities of MicroBooNE for high momentum exiting tracks.}

\begin{document}
\maketitle
\flushbottom

\section{Introduction and motivation}\label{sec:intro}

In this paper we summarize the theory of multiple Coulomb scattering (MCS) and describe how the underlying Highland formula is retuned based on Monte Carlo simulation for use in liquid-argon time-projection chambers (LArTPCs). We present a maximum likelihood based algorithm that is used to determine the momentum of particles in a LArTPC. The only way to determine the momentum of a particle that exits the active volume of a LArTPC is through MCS measurements. We demonstrate that this technique works well for a sample of fully contained muons from Booster Neutrino Beam (BNB) $\nu_\mu$ charged-current (CC) interactions, and determine the resolutions and biases of the measurement. In addition we demonstrate the performance of the method on simulated exiting tracks.\\

MicroBooNE (Micro Booster Neutrino Experiment) is an experiment that uses a large LArTPC to investigate the excess of low energy events observed by the MiniBooNE experiment \cite{Aguilar-Arevalo:2013pmq} and to study neutrino-argon cross-sections. MicroBooNE is the first detector of the Short-Baseline Neutrino (SBN) \cite{SBNwhitepaper} physics program at the Fermi National Accelerator Laboratory (Fermilab), to be joined by two other LArTPCs: the Short Baseline Near Detector (SBND) and the Imaging Cosmic And Rare Underground Signal (ICARUS) detector \cite{ICARUS_maincitation}. In addition to producing valuable physics output, MicroBooNE serves as an important source of detector and reconstruction development for future LArTPC experiments, such as the Deep Underground Neutrino Experiment (DUNE) \cite{DUNE_citation}.\\

The MicroBooNE detector \cite{ub_detectorpaper} consists of a rectangular time-projection chamber (TPC) with dimensions 2.6 m $\times$ 2.3 m $\times$ 10.4 m (width $\times$ height $\times$ length) located 470 m downstream from the Booster Neutrino Beam (BNB) target \cite{BNB_citation}. LArTPCs allow for precise three-dimensional reconstruction of particle interactions. For later reference, the $z$ axis of the detector is horizontal, along the direction of the BNB, while the $x$ direction of the TPC corresponds to the drift coordinate and the $y$ direction is the vertical direction. The mass of active liquid argon contained within the MicroBooNE TPC volume is about 90 tons, out of a total mass of 170 tons.\\

\begin{figure}[ht!]
\centering
	\includegraphics[width=0.9\textwidth]{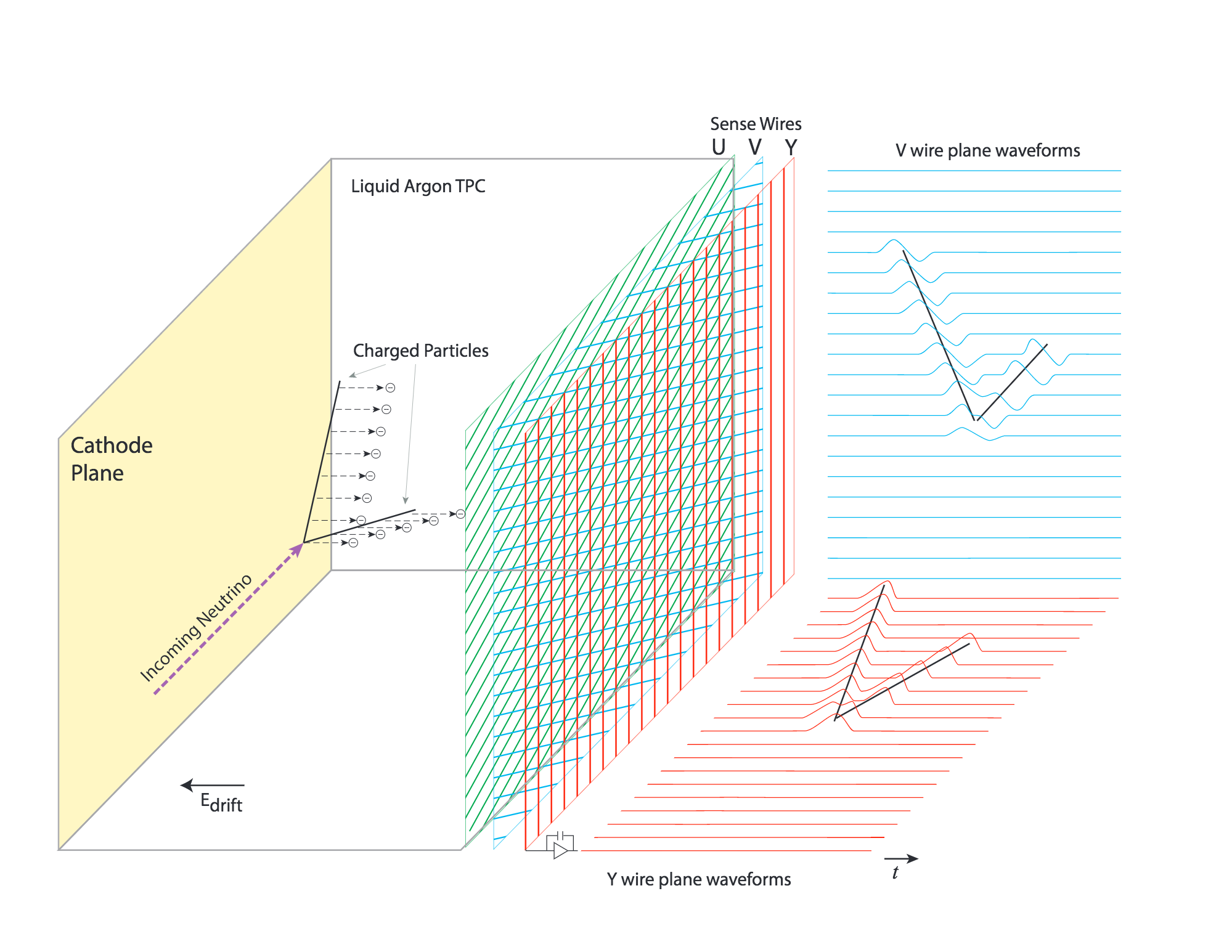} \\
\caption{A diagram of the time projection chamber of the MicroBooNE detector \cite{ub_detectorpaper}. PMTs (not shown) are located behind the wire planes.}\label{detector_fig}
\end{figure}

A set of 32 photomultiplier tubes (PMTs) and three planes of TPC wires with 3 mm spacing at angles of 0, and $\pm$ 60 degrees with respect to the vertical are used for event reconstruction. The cathode plane operating voltage is -70 kV. As illustrated in figure \ref{detector_fig}, a neutrino in the beam interacts with an argon nucleus and the charged outgoing particles traverse the medium, lose energy and leave an ionization trail. The resulting ionization electrons drift in a 273 $\text{V/cm}$ electric field to the wire planes constituting the anode. The passage of these electrons through the first two wire planes induces a signal in the wires, and their collection on the third plane also generates a signal. These signals are used to create three distinct two-dimensional views (in terms of wire and time) of the event. Combining these wire signals allow for full three-dimensional reconstruction of the event, with PMT signals providing information about the absolute drift ($x$) coordinate. The boundaries of the fiducial volume used in this analysis are set back from the six faces of the active volume by distances of between 20 and 37 cm, depending on the face, to reduce the impact of electric-field non-uniformities near the edges of the TPC. This volume corresponds to a mass of 55 tons.\\

The Booster Neutrino Beam (BNB) is composed predominantly of muon neutrinos ($\nu_\mu$) with a peak neutrino energy of about 0.7 GeV. Some of these neutrinos undergo charged current ($\nu_\mu$CC) interactions in the TPC and produce muons and other particles. For muon tracks that are completely contained in the TPC, we calculate the momentum with a measurement of the length of the particle's track, or with calorimetric measurements which come from wire signal size measurements. Roughly half of the muons from BNB $\nu_\mu$CC interactions in MicroBooNE are not fully contained in the TPC, and therefore using an established length-based or calorimetry-based method to determine the momenta for these uncontained tracks is not a possibility; the only way to determine their momenta is through MCS. \\

\section{Multiple Coulomb scattering}

\begin{figure}[ht!]
\centering
	\includegraphics[width=0.9\textwidth]{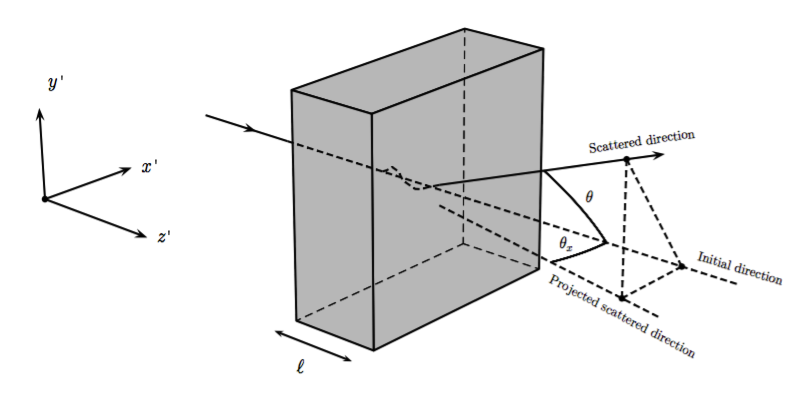} \\
\caption{The particle's trajectory is deflected as it traverses the material. The angular scatter in the labeled $x'$ direction is shown as $\theta_x$.}\label{mcs_nocap_fig}
\end{figure}

Multiple Coulomb scattering occurs when a charged particle traverses a medium and undergoes electromagnetic scattering off atomic nuclei. This scattering perturbs the original trajectory of the particle within the material (figure \ref{mcs_nocap_fig}). For a given initial momentum $p$, the angular deflection scatters of a particle in either the $x'$ direction or $y'$ direction (as indicated in the aforementioned figure) are modeled with a Gaussian distribution centered at zero with an RMS width, $\sigma_o^{\text{HL}}$, given by the Highland formula \cite{highland,highland-lynch-dahl} 

\begin{equation}\label{highland_eqtn}
	\sigma_o^{\text{HL}}=\frac{S_2}{p\beta c}z\sqrt{\frac{\ell}{X_0}}\left[1+\epsilon\times\ln\left(\frac{\ell}{X_0}\right)\right],
\end{equation}

\noindent where $\beta$ is the ratio of the particle's velocity to the speed of light (assuming the particle is a muon), $\ell$ is the distance traveled inside the material, $z$ is the magnitude of the charge of the particle (unity, for the case of muons), and $X_0$ is the radiation length of the target material (taken to be a constant $14$~cm in liquid argon). $S_2$ and $\epsilon$ are parameters determined to be 13.6 MeV and 0.038, respectively. So called ``mixture models''~\cite{mixmodel1,mixmodel2} which model both the core and tails of scattering distributions are not used in this study, though their inclusion may potentially improve algorithm performance. In this study, a modified version of the Highland formula is used that includes a detector-inherent angular resolution term, $\sigma_o^{\text{res}}$
\begin{equation}\label{modified_highland_eqtn}
\sigma_{o} = \sqrt{ (\sigma_o^{\text{HL}})^2 + (\sigma_o^{\text{res}})^2}.
\end{equation}
For this analysis, the $\sigma_o^{\text{res}}$ term is given a fixed value of 3 mrad which has been determined to be an acceptable value based on MicroBooNE simulation studies of muons at higher momenta. At 4.5 GeV/c muon momentum and $l\approx X_0$, equation \ref{highland_eqtn} predicts an RMS angular scatter of 3 mrad, comparable to the detector resolution. The fully contained muons addressed in this analysis have momenta below $1.5$~$\text{GeV/c}$, making the impact of this detector resolution minimal for that sample.\\

With the Highland formula, the momentum of a track-like particle can be determined using only the 3D reconstructed track information, without any calorimetric or track range information. In neutrino physics experiments, emulsion detectors like those employed by the DONUT \cite{DONUT_paper} and OPERA \cite{OPERA_paper} collaborations have used MCS to determine particle momenta. Additionally, the MACRO \cite{MACRO_paper} collaboration at Gran Sasso Laboratory utilized this technique. For LArTPCs, the ICARUS collaboration has described the MCS-based determination of particle momentum using a variety of methods \cite{icarus_mcs_paper,new_icarus_paper}. The likelihood-based method discussed in this paper for use in the MicrobooNE detector and described in detail in section \ref{MCS_technique_section}, has improved on the ICARUS method by tuning the underlying phenomenological formula.

\subsection{Tuning the Highland formula for argon}\label{highland_tuning_section}

The Highland formula as written in equation \ref{highland_eqtn} originates from a 1991 publication by G. R. Lynch and O. I. Dahl \cite{highland-lynch-dahl}. The parameters in the equation ($S_2$ and $\epsilon$) were determined using a global fit to MCS simulated data using a modified GEANT simulation package of 14 different elements and 7 thickness ranges. All of the simulated particles were relativistic, with $\beta=1$. The materials studied ranged from hydrogen (with Z=1) to uranium (with Z=92). Given that the parameters in the formula were determined from a single fit to a wide range of Z with a wide range of material thicknesses, there is reason to believe that these parameters could differ for scattering specifically in liquid argon with $l \approx X_0$. There is also reason to believe that these parameters might be momentum-dependent for particles with $\beta < 1$, which is the case for some of the contained muons in this analysis.\\

In order to re-tune these parameters for liquid argon, a large sample of muons are simulated with GEANT4\footnote{The GEANT4 version used in this simulation is 4.9.6.} \cite{GEANT4_citation} in the MicroBooNE TPC and their true angular scatters are used in a fit, with $l = X_0$. The reason for using $l = X_0$ is that the Highland formula simplifies to remove its dependence on $\epsilon$

\begin{equation}\label{highland_simplified}
	\sigma_o^{\text{HL}}=\frac{S_2}{p\beta c}.
\end{equation}

\begin{figure}[ht!]
\begin{center}
\includegraphics[width=0.9\textwidth]{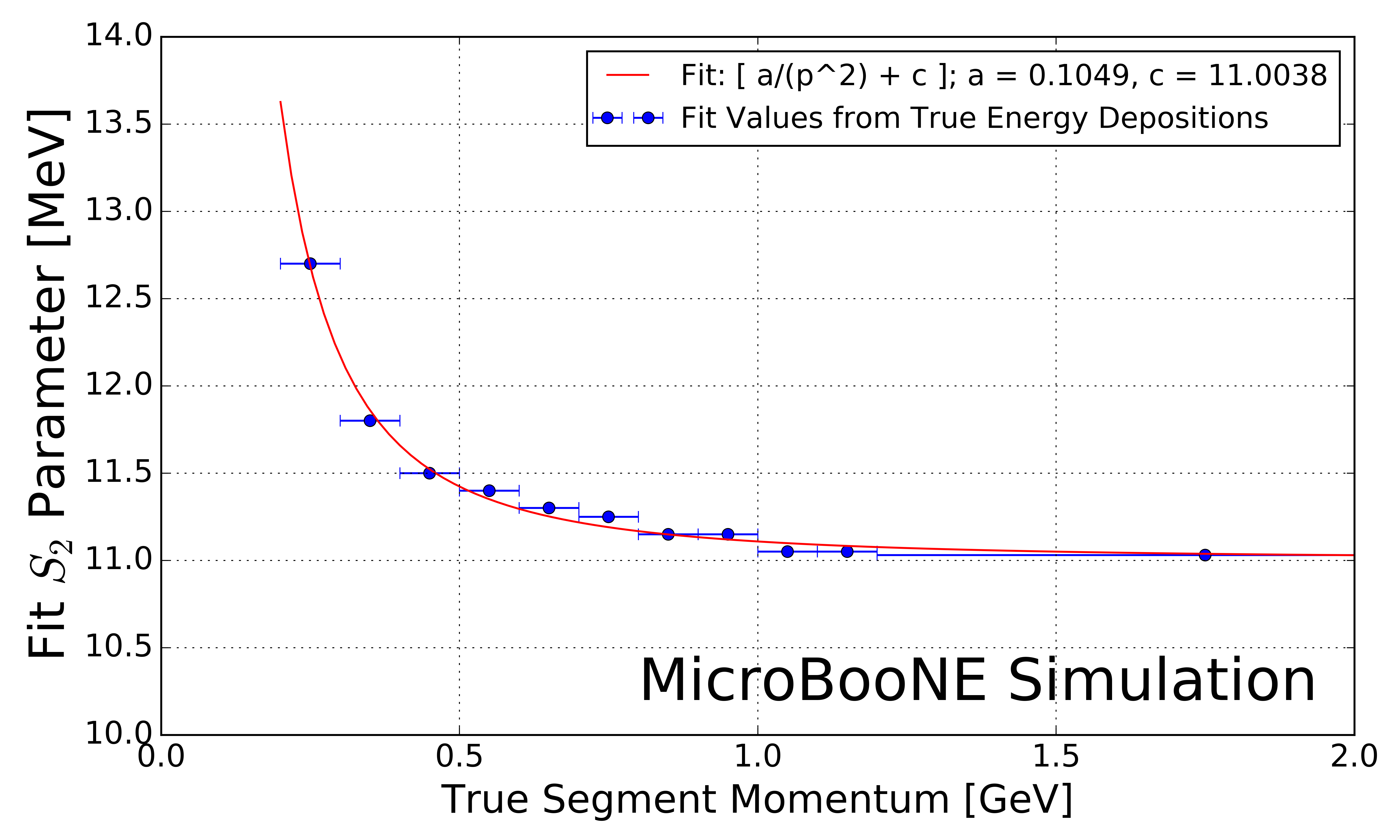}
\end{center}
\caption{Fitted Highland parameter $S_2$ as a function of true segment momentum for ${\ell} = X_0$ simulated muons in the MicroBooNE LArTPC. Blue x- error bars indicate the true momentum bin width with data points drawn at the center of each bin. Shown in red is a fit to these data points with functional form $a\times (p(\text{GeV}))^{-2} + c$, with best fit values for parameters $a$ and $c$ shown in the legend.}
\label{retune_highland_fig1}
\end{figure}

The $S_2$ parameter in equation \ref{highland_simplified} is fit for as a function of true muon momentum at each scatter, in order to explore the $\beta$ dependence of this parameter. Note that to use a segment length other than $X_0$ a simultaneous fit both for $S_2$ and for $\epsilon$ is necessary, which is not described in this paper. The fitted $S_2$ parameter value as a function of true momentum is shown in figure \ref{retune_highland_fig1}.


The fitted value of $S_2$ is always less than the nominal $13.6\text{ MeV}$ for momentum greater than $0.25\text{ GeV/c}$ and asymptotically approaches a constant at higher momentum (where $\beta = 1$) of about $11.0\text{ MeV}$. The value increases in the momentum region where $\beta < 1$. Shown in red is a fit to these data points with functional form $a\times (p(\text{GeV}))^{-2} + c$, with best fit values for floating parameters $a$ and $c$ being $0.105~\text{MeV}$ and $11.004\text{ MeV}$ respectively. This functional form is chosen because it captures the trend in the fit value of $S_2$ with respect to momentum, and asymptotically approaches a constant value when $\beta$ approaches 1. This function, used as a replacement for the $S_2$ parameter in the Highland formula, will henceforth be referred to as $\kappa(p)$:
\begin{equation}\label{kappa_equation}
\kappa(p) = \frac{0.105 \text{ MeV}}{(p(\text{GeV}))^2} + 11.004\text{ MeV}.
\end{equation}

\begin{figure}[ht!]
\begin{center}
\includegraphics[width=0.9\textwidth]{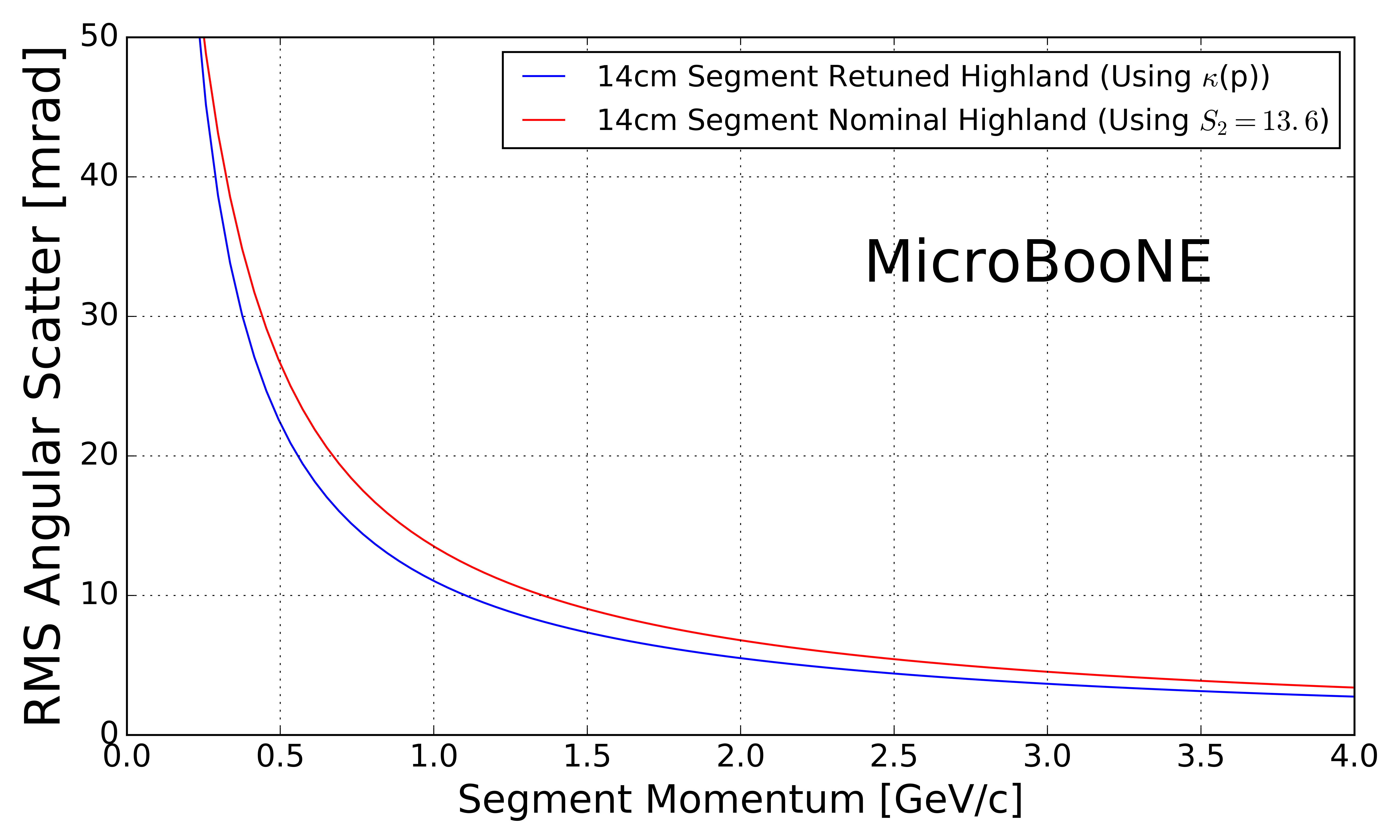}
\end{center}
\caption{The Highland scattering RMS $\sigma_o^{\text{HL}}$ for 14 cm segment lengths and $\sigma_o^{\text{res}}=0$ as a function of true momentum before and after tuning. In red is shown equation \ref{highland_simplified} (the nominal Highland formula using $S_2 = 13.6~\text{MeV}$) and in blue is the retuned Highland formula (replacing $S_2$ with $\kappa(p)$).}
\label{retune_highland_fig2}
\end{figure}

To visualize the Highland formula for ${\ell} = X_0$ both before and after the $\kappa(p)$ replacement, see figure \ref{retune_highland_fig2}. It is recommended that future LArTPC experiments use this parameterization of the Highland formula, or at the very least conduct their own studies to tune the Highland formula for scattering in argon. This formulation can also be checked in LAr-based test-beam experiments such as LArIAT \cite{LARIAT_citation}.\\

With ${\ell} = X_0$, the form of the Highland equation used in this analysis is therefore
\begin{equation}\label{modified_highland_eqtn_kappa}
\sigma_{o}^{\text{RMS}} = \sqrt{ (\sigma_o)^2 + (\sigma_o^{\text{res}})^2} = \sqrt{ \left(\frac{\kappa(p)}{p\beta c}\right)^2 + (\sigma_o^{\text{res}})^2 }.
\end{equation}

\section{MCS implementation using the maximum likelihood method}\label{MCS_technique_section}

This section explains in detail how the phenomenon of multiple Coulomb scattering is used to determine the momentum of a muon track reconstructed in a LArTPC. In general, the approach is as follows:
\begin{enumerate}
\item The three-dimensional track is divided into segments of configurable length.
\item The scattering angles between consecutive segments are measured.
\item Those angles combined with the modified, tuned Highland formula (equation \ref{modified_highland_eqtn_kappa}) are used to build a likelihood that the particle has a specific momentum, taking into account energy loss in upstream segments of the track.
\item The momentum corresponding to the maximum of the likelihood is chosen to be the MCS-computed momentum.
\end{enumerate}
Each of these steps is discussed in detail in the following subsections.\\


\subsection{Track segmentation and scattering angle computation}\label{track_segmentation_and_scattering_angle_computation_section}

Track segmentation refers to the subdivision of three-dimensional reconstructed trajectory points of a reconstructed track into portions of definite length. In this analysis, the tracks are automatically reconstructed by a projection matching algorithm \cite{Antonello:2012hu} run on the output of MicroBooNE's Pandora-based neutrino event reconstruction chain \cite{Marshall:2015rfa}. The algorithm constructs the three-dimensional trajectory points by combining two-dimensional hits reconstructed from signals on the different wire planes along with timing information from the photomultiplier tubes. The segmentation process begins at the start of the track, and iterates through the trajectory points in order, defining segment start and stop points based on the straight-line distance between them. There is no overlap between segments. Given the subset of the three-dimensional trajectory points that corresponds to one segment of the track, a three-dimensional linear fit is applied to the data points, weighting all trajectory points equally in the fit. In this analysis, a segment length of 14 cm is used, which is a tunable parameter that has been chosen as described in the derivation of $\kappa(p)$ (equation \ref{kappa_equation}). While the optimal segment length may vary as a function of particle momentum, static segment lengths are used in this analysis as is required by the specific retuning of the Highland formula described in section \ref{highland_tuning_section}.\\

With the segments defined, the scattering angles between the linear fits from adjacent segments are computed. A coordinate transformation is performed such that the $z'$ direction is oriented along the direction of the linear fit to the first of the segment pair. The $x'$ and $y'$ coordinates are chosen such that all of $x'$, $y'$, and $z'$ are mutually orthogonal and right-handed, as shown in figure \ref{mcs_nocap_fig}. The scattering angles with respect to the $x'$ direction and the $y'$ direction are computed as input to the MCS algorithm. Only the scattering angle with respect to the $x'$ direction is drawn in figure \ref{mcs_nocap_fig}.

\subsection{Maximum likelihood theory}\label{likelihood_theory_section}

The normal probability distribution for a scattering angle in either the $x'$ or $y'$ direction, $\Delta\theta$, with an expected Gaussian uncertainty $\sigma_o$ and mean of zero is given by
\begin{equation}
f_X(\Delta\theta) = (2\pi\sigma_o^2)^{-\frac{1}{2}}\exp\left[-\frac{1}{2}\left(\frac{\Delta\theta}{\sigma_o}\right)^2\right].
\end{equation}

Here, $\sigma_o$ is the RMS angular deflection computed by the modified, tuned Highland formula (equation \ref{modified_highland_eqtn_kappa}), which is a function of the momentum and the length of that segment. Since energy is lost between segments along the track, $\sigma_o$ increases for each angular measurement along the track. We therefore replace $\sigma_o$ with $\sigma_{o,j}$, where $j$ is an index representative of the segment. \newline

To obtain the likelihood, we take the product of $f_X(\Delta\theta_j)$ over all $n$ of the $\Delta\theta_j$ segment-to-segment scatters along the track. This product can be written as
\begin{equation}
L(\sigma_{o,1},...,\sigma_{o,n};\Delta\theta_1,...,\Delta\theta_n) = (2\pi)^{-\frac{n}{2}}\times\prod_{j=1}^{n}(\sigma_{o,j})^{-1} \times \exp\left[-\frac{1}{2}\sum_{j=1}^{n}\left(\frac{\Delta\theta_j}{\sigma_{o,j}}\right)^2\right].
\end{equation}

Rather than maximizing the likelihood it is more computationally convenient to instead minimize the negative log likelihood. Inverting the sign and taking $\ln(L)$ gives an expression that is related to a $\chi^2$ variable:
\begin{equation}\label{leo_llhd_eqtn}
-l(\sigma_{o,1},...,\sigma_{o,n};\Delta\theta_1,...,\Delta\theta_n) = -\ln(L) = \frac{n}{2}\ln(2\pi) + \sum_{j=1}^{n}\ln(\sigma_{o,j}) + \frac{1}{2}\sum_{j=1}^{n}\left(\frac{\Delta\theta_j}{\sigma_{o,j}}\right)^2.
\end{equation}



\subsection{Maximum likelihood implementation}\label{maximum_likelihood_section}

Given a set of angular deflections in the $x'$ and $y'$ directions for each segment as described in section \ref{track_segmentation_and_scattering_angle_computation_section} a scan is done over the postulated initial energy, $E_t$, in steps of 1 MeV up to 7.5 GeV. The step with the smallest negative log likelihood (equation \ref{leo_llhd_eqtn}) is chosen as the MCS energy. Equation \ref{leo_llhd_eqtn} includes a $\sigma_{o,j}$ term that changes for consecutive segments because their associated energy is decreasing. The energy of the $j$th segment is related to $E_t$ by

\begin{equation}\label{segment_E_equation}
E_{j} = E_t - \Delta E_{j},
\end{equation}

\noindent where $\Delta E_{j}$ is the energy loss upstream of this segment, computed by integrating the muon stopping power curve given by the Bethe-Bloch equation described by the Particle Data Group (PDG) \cite{stoppingpowersource} along the length of track upstream of this segment. Equation \ref{segment_E_equation} introduces a minimum allowable track energy determined by the range of the track, as $E_{j}$ must remain positive. The use of the Bethe-Bloch equation to determine $\Delta E_{j}$ impacts the MCS algorithm resolution for fully contained tracks, but does not for exiting tracks where much of the ionization energy loss is not visible. This value of segment energy, $E_{j}$, is converted to a momentum $p$ with the relativistic energy-momentum relation assuming the muon mass, and is then used to predict the RMS angular scatter for that segment ($\sigma_o$) by way of equation \ref{modified_highland_eqtn_kappa}. 

\section{Range-based energy validation from simulation}\label{Range_Energy_Validation_section}
\begin{figure}
\centering
\includegraphics[width=0.9\textwidth]
	{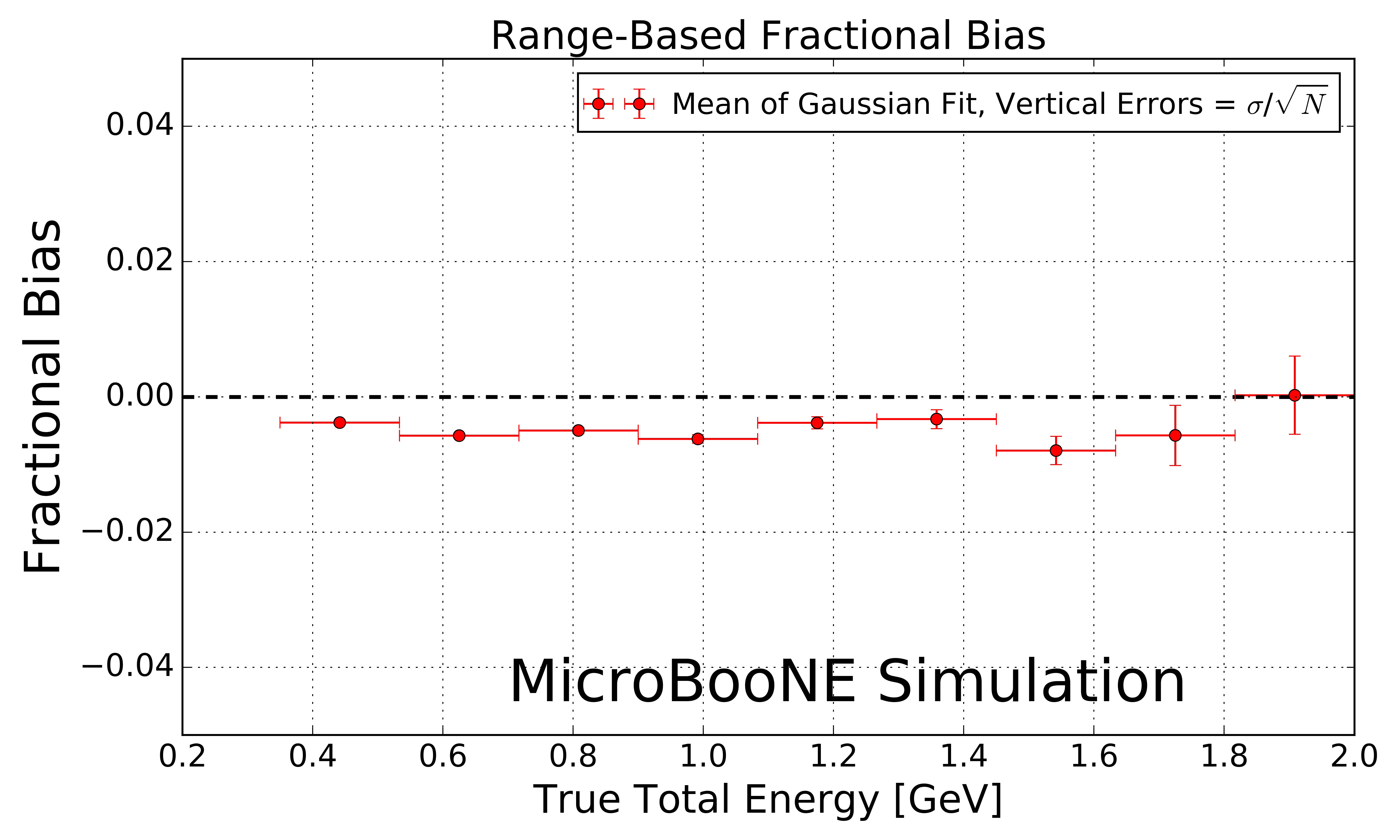}
\includegraphics[width=0.9\textwidth]
	{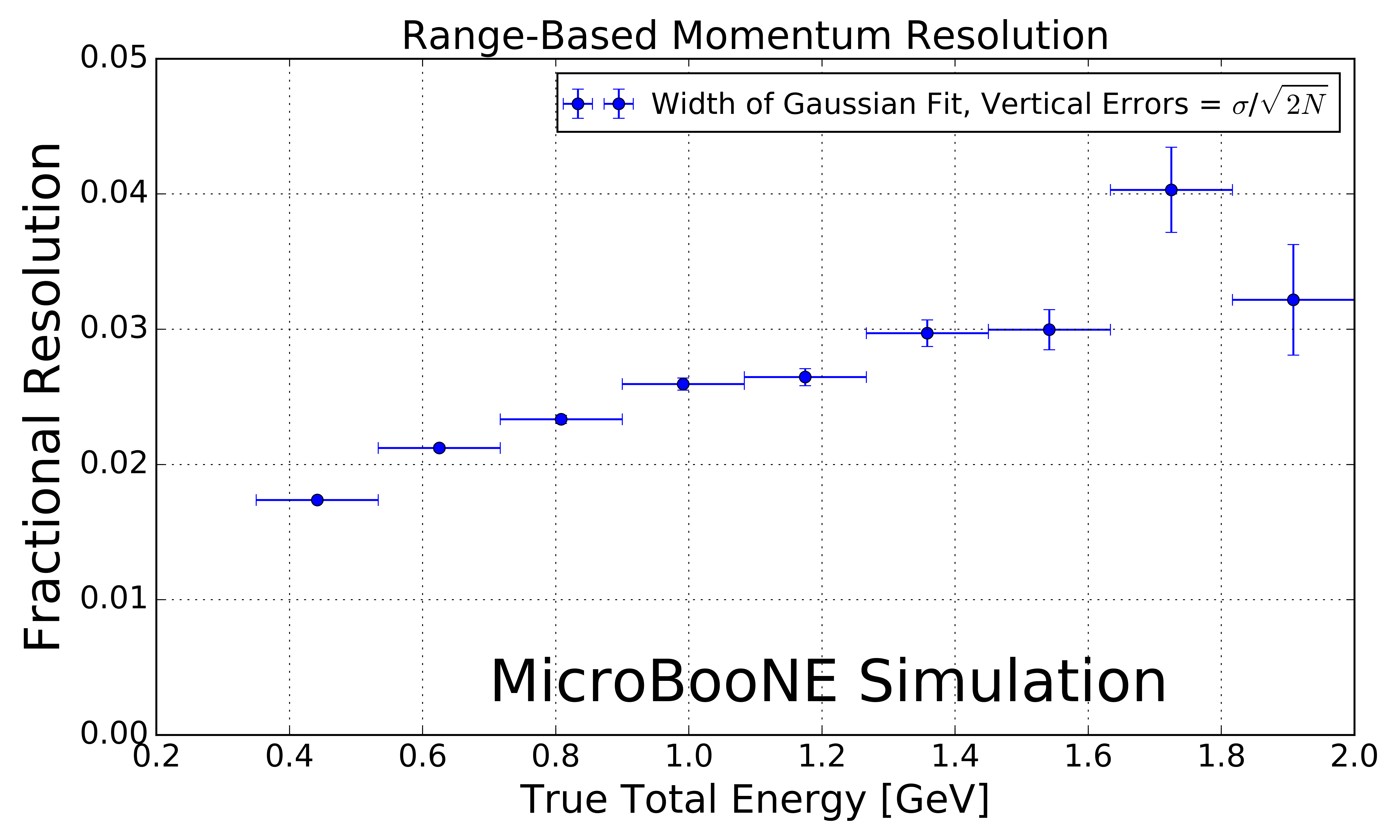}
\caption{Range-based energy fractional bias (top) and resolution (bottom) from a sample of simulated fully contained BNB neutrino-induced muons using true starting and stopping positions of the track. The bias is less than 1\% and the resolution is better than 4\%.}
\label{true_range_bias_resolution_MCTrack_fig}
\end{figure}

In order to quantify the performance of the MCS energy estimation method on fully contained muons in data, an independent determination of energy is needed. Range-based energy, $E_{\text{range}}$ is used here because the true energy $E_{\text{true}}$ will not be known in analyzing detector data. The stopping power of muons in liquid argon is well described by the continuous slowing-down approximation (CSDA) by the Particle Data Group, and agrees with data at the sub-percent level \cite{MIPenergysource,PDG_spline_table,NISTdata}. By using a linear interpolation between points in the stopping power table of ref. \cite{PDG_spline_table}, the length of a track can be used to reconstruct the muon's total energy with good accuracy. A simulated sample of fully contained BNB neutrino-induced muons longer than one meter is used to quantify the bias and resolution for the range-based energy estimation technique. The range is defined as the straight-line distance between the true starting point and true stopping point of a muon, even though the trajectories are not perfectly straight lines. The bias and resolution are computed in bins of true total energy of the muons by performing a least-squares minimization Gaussian function fit to a distribution of the fractional energy difference $(E_{\text{Range}}-E_{\text{True}})/(E_{\text{True}})$ in each bin. The mean of each Gaussian yields the bias for that true energy bin, and the width indicates the resolution. The mean and standard deviation of the raw data agrees with the mean and width (respectively) of the Gaussian fits for these fully contained muons. Figure \ref{true_range_bias_resolution_MCTrack_fig} shows the bias and resolution for the range-based energy reconstruction method. The bias is less than 1\% and the resolution for this method of energy reconstruction increases slightly with true muon energy but remains on the order of (2-4)\%. This result demonstrates that range-based energy (and therefore range-based momentum) is a good estimator of the true energy (momentum) of a reconstructed contained muon track in data, assuming that the track is well reconstructed in terms of length.

\section{MCS performance on beam neutrino-induced muons in MicroBooNE data}\label{data_performance_section}

\subsection{Input sample}\label{input_sample_section}
This part of the analysis is based on triggered neutrino interaction events in MicroBooNE data corresponding to $\approx 5 \times 10^{19}$ protons on target, which is a small subset ($<$10\%) of the nominal protons on target already delivered to the detector. These events are run through a fully automated reconstruction chain that produces reconstructed objects including three-dimensional neutrino interaction points (vertices), three-dimensional tracks (as described in section \ref{track_segmentation_and_scattering_angle_computation_section}) for each outgoing secondary particle from the interaction, and PMT-reconstructed optical flashes from the interaction scintillation light. The fiducial volume used in this analysis is defined in section \ref{sec:intro}.

\subsection{Event selection}
The following selection criteria are placed on the reconstructed objects to select $\nu_\mu$ charged-current interactions in which a candidate muon track exiting the interaction vertex is fully contained within the fiducial volume:
\begin{enumerate}
\item The event must have at least one bright optical flash, reconstructed from PMT timing signals, in coincidence with the expected BNB-neutrino arrival time.
\item Two or more reconstructed tracks must originate from the same reconstructed vertex within the fiducial volume.
\item The $z$ coordinate of the optical flash, as determined by the pulse height and timing of signals in the 32 PMTs, must be within 70 cm of any point on the $z$ projection of the candidate muon track.
\item For events with exactly two tracks originating from the vertex, additional calorimetric criteria are applied to mitigate backgrounds from cosmic muons that arrive in time with the passage of the beam, then stop and decay to an electron that is reconstructed as a track.
\item The longest track originating from the vertex is assumed to be a muon, and it must be fully contained within the fiducial volume.
\item The length of the longest track must be $>$1~m in order to have sufficient sampling points in the MCS likelihood to obtain a reasonable estimate of momentum.
\end{enumerate}

These selection criteria are chosen to select a sample of tracks with high purity. In this sample of MicroBooNE data, 598 events (tracks) remain after all selections. The low statistics in this sample is due to the size of the input sample and the low efficiency associated with the applied high-purity selection, described in section \ref{input_sample_section}. Each of these events (tracks) was scanned by hand with a 2D interactive event display showing the raw wire signals of the interaction from each wire plane, with the 2D projection of the reconstructed muon track and vertex overlaid. The scanning was done to ensure the track is well reconstructed with start point close to the reconstructed vertex and end point close to the end of the visible wire-signal track in all three planes. During the scanning, obvious mis-identification topologies were removed. An example of such a topology is a stopping cosmic-ray muon decaying into an electron. After rejecting events (tracks) based on hand scanning, 396 tracks remain for analysis.

\subsection{Validation of the Highland formula}\label{highland_validation_section}
\begin{figure}[ht!]
\centering
	\includegraphics[width=0.9\textwidth]{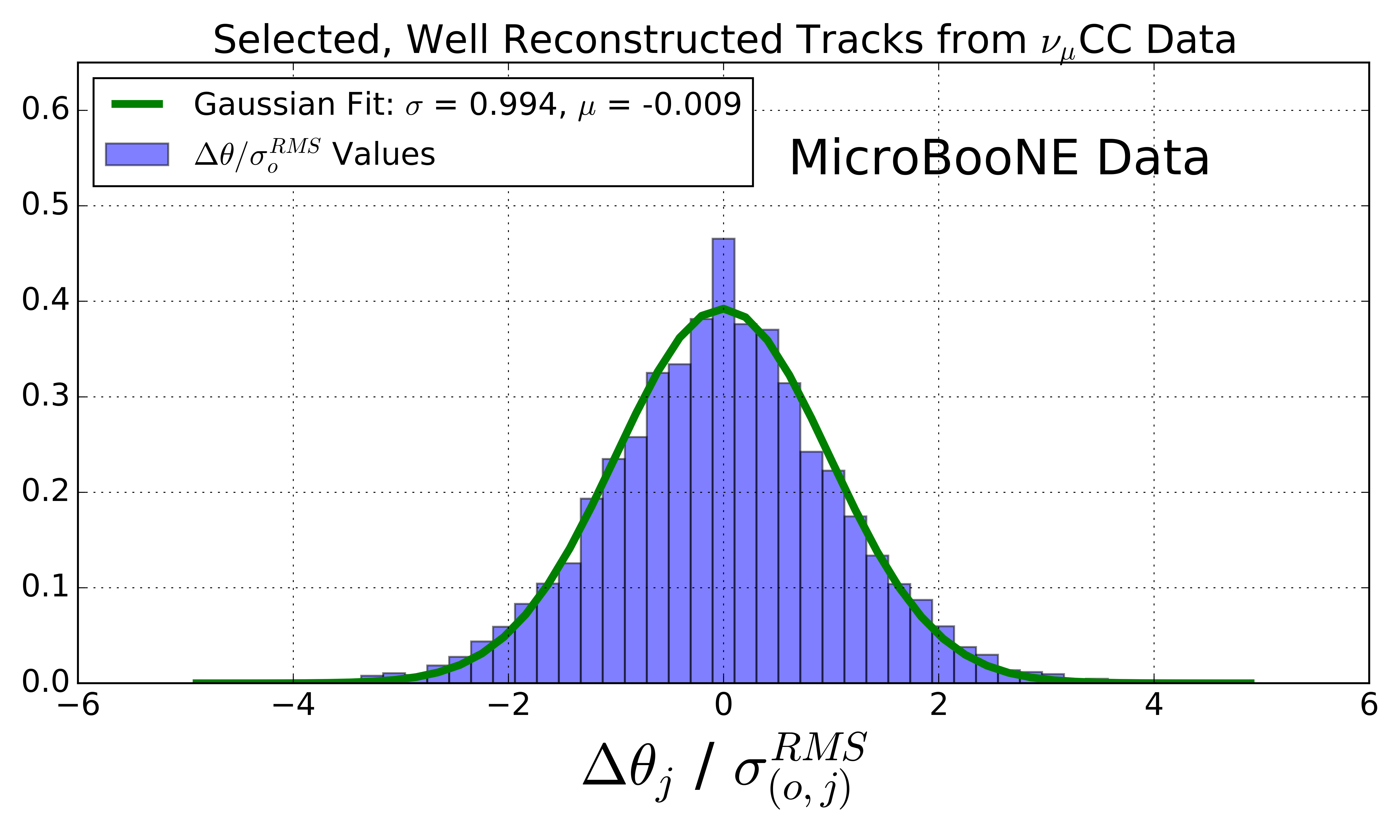} \\
\caption{Segment-to-segment measured angular scatters in both the $x'$ and $y'$ directions divided by the width $\sigma_o^{\text{RMS}}$ predicted by the Highland formula (equation \ref{highland_eqtn}) for the automatically selected beam neutrino-induced fully contained muon sample in MicroBooNE data after hand scanning to remove poorly reconstructed tracks and obvious mis-identification topologies.}\label{Highland_validation_fig}
\end{figure}

The Highland formula indicates that distributions of angular deviations of the track, segment by segment, in both the $x'$ and $y'$ directions divided by the width predicted from the Highland equation $\sigma_o^{\text{RMS}}$ (equation \ref{modified_highland_eqtn_kappa}) should be Gaussian with a width of unity. In order to calculate the momentum $p$ in the Highland equation, $p$ for each segment is computed with equation \ref{segment_E_equation}, where $E_t$ comes from the converged MCS-computed momentum of the track. For each consecutive pair of segments in this sample of 396 tracks, the angular scatter divided by the Highland expected RMS (including detector resolution term, $\sigma_o^{\text{res}}$) is an entry in the area-normalized distribution shown in figure \ref{Highland_validation_fig}. These 396 tracks have on average 12 segments each, therefore this histogram has approximately $396\times12\times2=9504$ entries. The additional factor of 2 comes from angular scatters both in the $x'$ and $y'$ directions. The distribution has an RMS of unity, thus validating the MCS technique used in this analysis.

\subsection{MCS momentum validation}\label{MCS_Momentum_Validation_DataRecoTrack_section}

\begin{figure}[ht!]
\centering
	\includegraphics[width=0.9\textwidth]{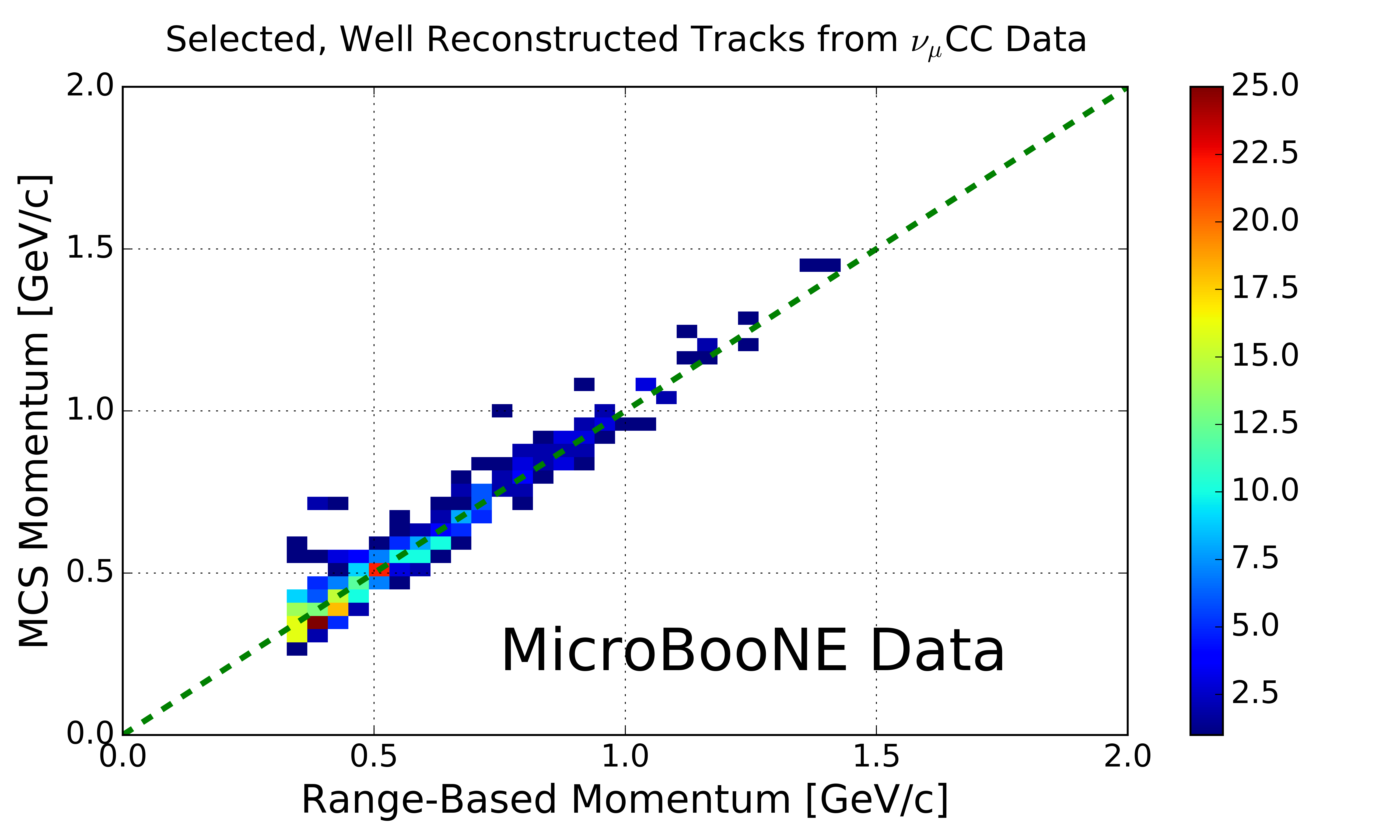} \\
\caption{MCS-computed momentum versus range momentum for the automatically selected beam neutrino-induced fully contained muon sample in MicroBooNE data after hand scanning to remove poorly reconstructed tracks and obvious mis-identification topologies. The color (z) scale indicates number of tracks.}\label{realdata_goodhandscan_fig}
\end{figure}

\begin{figure}
\centering
	\includegraphics[width=0.9\textwidth]{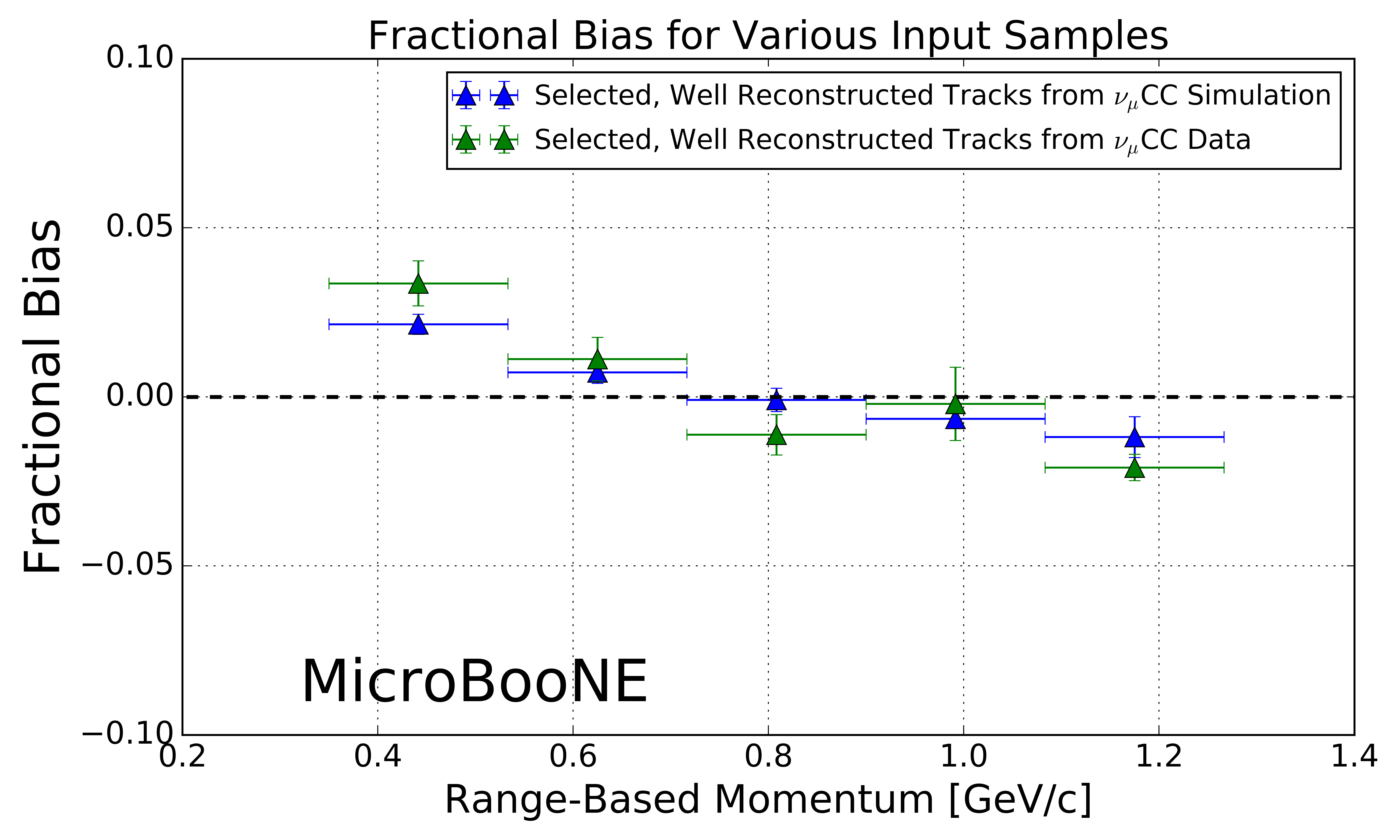}
	\includegraphics[width=0.9\textwidth]{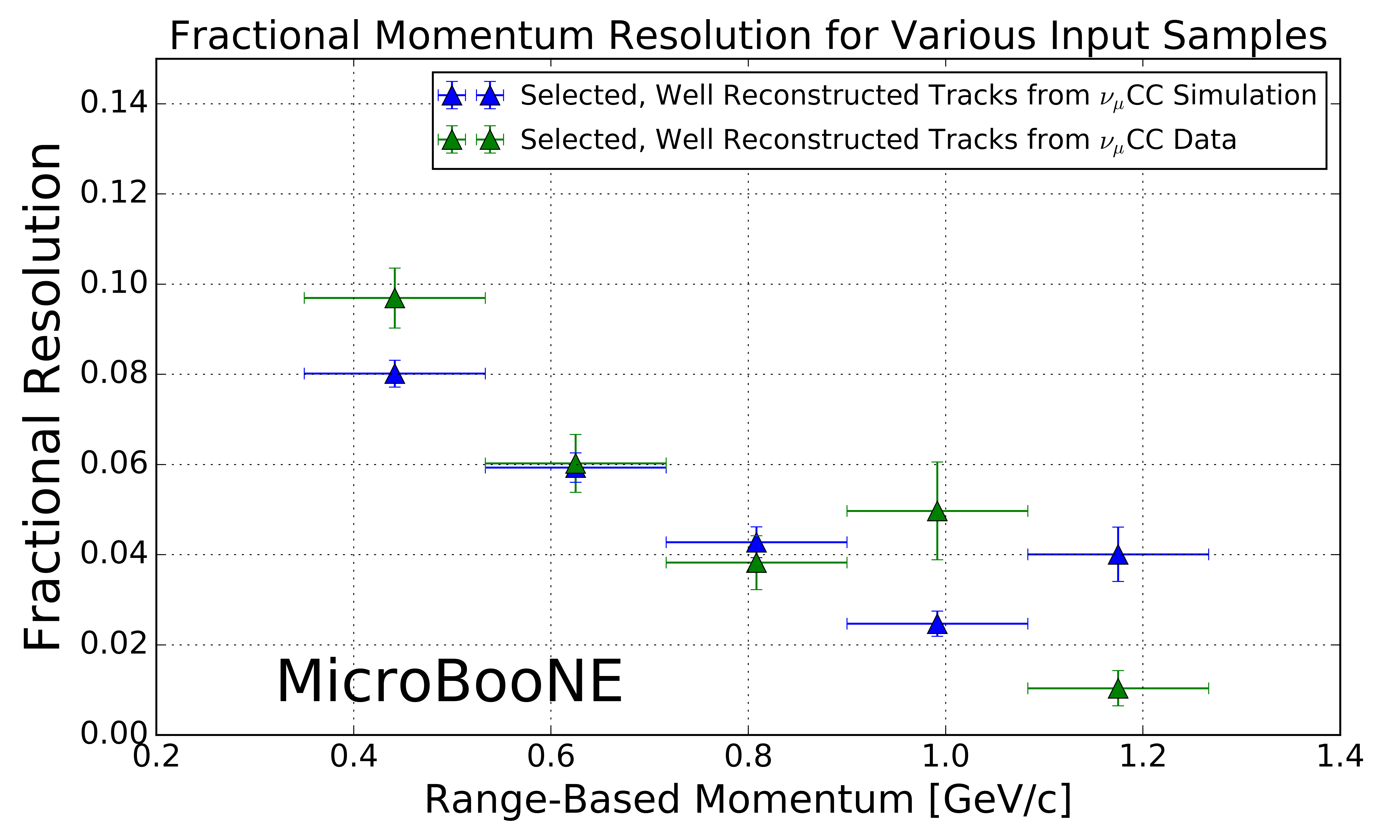}
\caption{Inverse momentum difference (as defined in the text) fractional bias (top) and resolution (bottom) for automatically selected contained $\nu_\mu$CC-induced muons from full simulated BNB events with cosmic overlay where the track matches with the true muon track (blue), and automatically selected and hand-scanned (see text) contained $\nu_\mu$CC-induced muons from MicroBooNE data (green).}\label{MCS_range_bias_resolution_DataRecoTrack_fig}
\end{figure}

MCS momentum versus range-based momentum for this sample of 396 tracks is shown in figure \ref{realdata_goodhandscan_fig}. The fractional bias and resolution as a function of range-based momentum for this sample is shown in figure \ref{MCS_range_bias_resolution_DataRecoTrack_fig}. In order to compute this bias and resolution, distributions of fractional inverse momentum difference $(p_{\text{MCS}}^{-1} - p_{\text{Range}}^{-1})/(p_{\text{Range}}^{-1})$ in bins of range-based momentum $p_{\text{Range}}$ are fit to Gaussian functions, where the mean of the fit determines the bias while the width of the fit determines the resolution for that bin. Inverse momentum is used here because the binned distributions are more Gaussian since the Highland formula measures inverse momentum in terms of track angles that have reasonably Gaussian errors. Simply using the mean and RMS of the binned distributions yields similar results. Also shown in this figure are the bias and resolutions for a simulated sample consisting of a full BNB simulation with CORSIKA-generated \cite{corsika_ref} cosmic overlays passed through an identical reconstruction and event selection chain. Rather than hand scanning this sample, true simulation information is used by requiring the longest reconstructed track to be matched well to the true starting and stopping point of the muon from the $\nu_\mu$CC interaction. This removes any mis-identifications or interference from the simulated cosmics. 

Figure \ref{MCS_range_bias_resolution_DataRecoTrack_fig} indicates a bias in the MCS momentum calculation on the order of a few percent, with a resolution that improves from about 10\% for contained reconstructed tracks in data and simulation with range momentum around 0.45 GeV/c (which corresponds to a length of about 1.5 m) to below 5\% for contained reconstructed tracks in data and simulation with range momentum about 1.15 GeV/c (which corresponds to a length of about 4.6 meters). Resolution improving with length of track is expected; the longer the track, the more angular scattering measurements can be made to improve the likelihood. The bias and resolutions show reasonable agreement between data and simulation.\\

\begin{figure}[ht!]
\centering
	\includegraphics[width=0.8\textwidth]{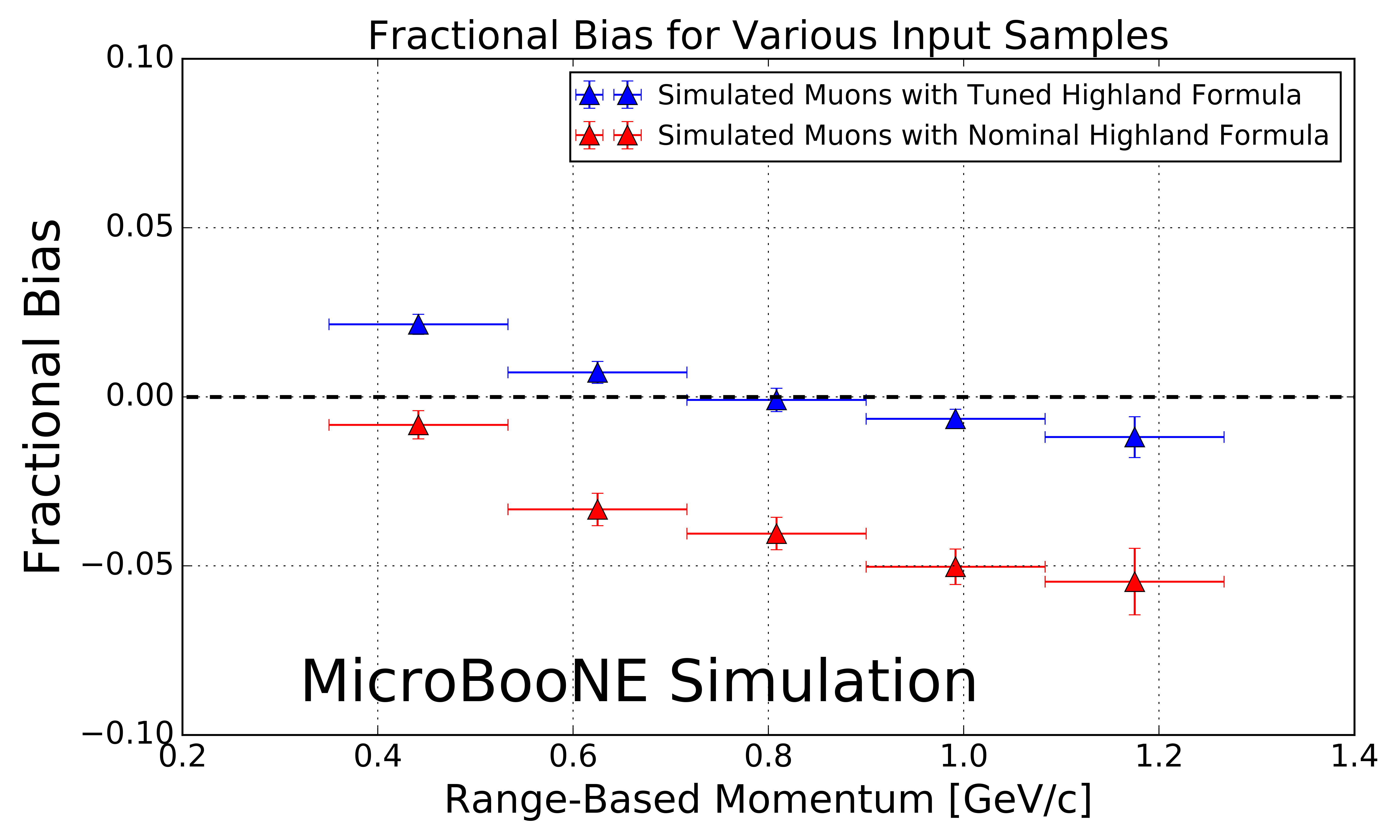}
	\includegraphics[width=0.8\textwidth]{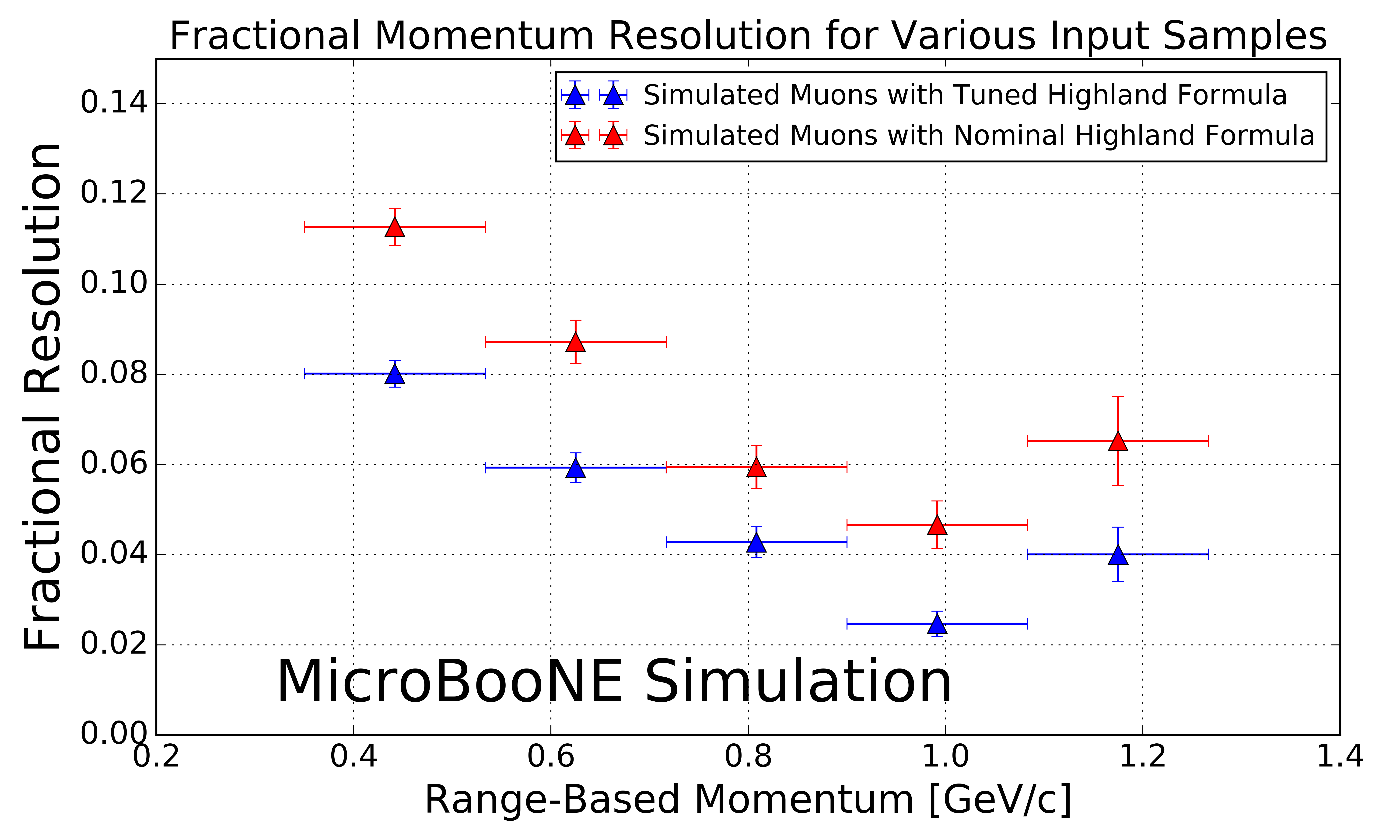}
\caption{Inverse momentum difference (as defined in the text) fractional bias (top) and resolution (bottom) for automatically selected contained $\nu_\mu$CC-induced muons from full simulated BNB events with cosmic overlay where the track matches with the true muon track both using the nominal Highland formula (equation \ref{modified_highland_eqtn}) (red) and the retuned Highland formula (equation \ref{modified_highland_eqtn_kappa}) (blue).}\label{MCS_range_bias_resolution_compareHighlandFormulas_fig}
\end{figure}

\subsection{Impact of Highland formula tuning}\label{highland_formula_tuning_impact_section}

In order to examine the impact of the Highland formula tuning described in section \ref{highland_tuning_section}, the fractional bias and resolution on the simulated sample of contained muons described in section \ref{MCS_Momentum_Validation_DataRecoTrack_section} both with the nominal Highland formula (equation \ref{modified_highland_eqtn}) and with the retuned Highland formula (equation \ref{modified_highland_eqtn_kappa}) are shown in figure \ref{MCS_range_bias_resolution_compareHighlandFormulas_fig}. Tuning the Highland formula improves the magnitude of the fractional bias to below 2\%, and improves the fractional resolution by 30 to 40\% of the untuned value.\\

\section{MCS performance on exiting muons from MicroBooNE simulation}
In this section we quantify the MCS algorithm performance for two samples of exiting muon tracks from simulated BNB $\nu_\mu$CC interactions within the MicroBooNE detector.  Simulated data must be used for this study since the range-based energy cannot be calculated for exiting tracks.  \\

The first sample of exiting tracks uses energy depositions within the TPC directly from the GEANT4 simulation and is therefore insensitive to reconstruction and detector effects.  This sample is referred to as the ``{\sc MCTracks}'' sample.   The {\sc MCTracks} sample does not depend on the details of the MicroBooNE detector response such as broken wires, noise, and TPC pulse digitization and also has no experimental measurement uncertainty included for the track segments.  The sample only includes depositions within the TPC active volume to mimic the track coverage for the exiting reconstructed track sample and all tracks used in the study are required to have a length of at least $1$~m within the active volume.   The study of the MCTrack sample, therefore, tests the inherent performance of the MCS algorithm to determine the momentum for exiting tracks and is not affected by the mis-reconstruction of the muon tracks and detector effects.\\

\begin{figure}[ht!]
\centering
	\includegraphics[width=0.9\textwidth]{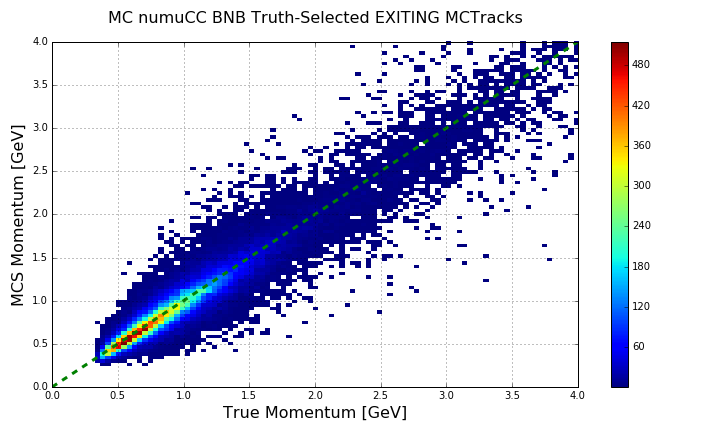} \\
\caption{MCS-computed momentum versus true momentum for the {\sc MCTracks} (truth-based) sample of simulated exiting muons from BNB $\nu_\mu$CC interactions in MicroBooNE with at least one meter of track contained within the TPC. The color (z) scale indicates number of tracks.}\label{MCS_true_comparison_exiting_MCTracks_fig}
\end{figure}

\begin{figure}
\centering
\includegraphics[width=0.45\textwidth]
	{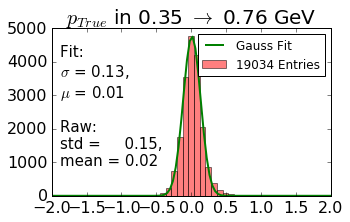}
\includegraphics[width=0.45\textwidth]
	{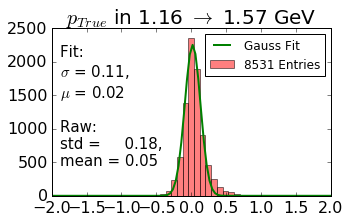}
\includegraphics[width=0.45\textwidth]
	{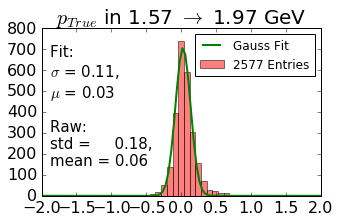}
\includegraphics[width=0.45\textwidth]
	{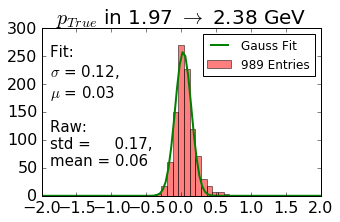}
\includegraphics[width=0.45\textwidth]
	{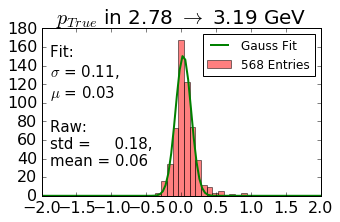}
\includegraphics[width=0.45\textwidth]
	{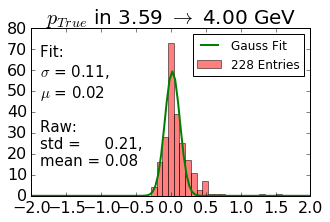}
\caption{Fractional momentum difference for six representative bins of true momentum for the MCTrack 
sample of simulated exiting muon tracks, which use the true GEANT hits unaffected by reconstruction effects. 
The y-axis is the number of tracks, and the x-axis is $(p_{\text{MCS}}^{-1} - p_{\text{true}}^{-1})/(p_{\text{true}}^{-1})$.}
\label{MCS_true_exiting_resolution_MCTracks_slices_fig}
\end{figure}

\begin{figure}
\centering
\includegraphics[width=0.9\textwidth]
	{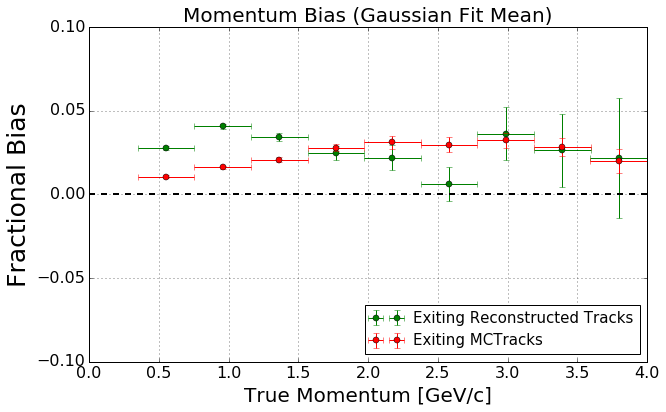}
\includegraphics[width=0.9\textwidth]
	{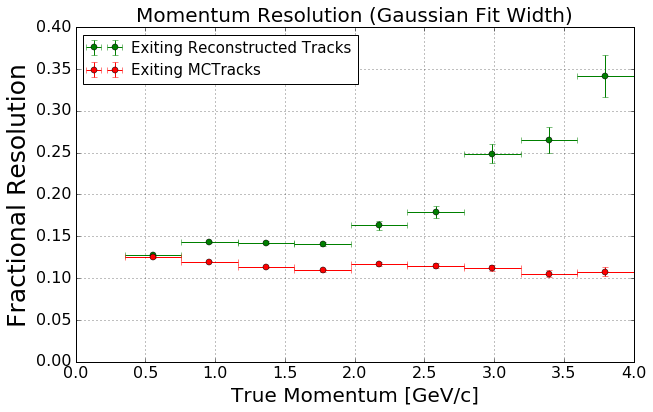}
\caption{MCS momentum fractional bias (top) and resolution (bottom) from Gaussian fits to the reconstructed momentum as a function of true momentum for the exiting muon using the {\sc MCTracks} (truth-based) sample and using the fully-simulated sample.} 
\label{exitingMCTrack_and_RecoTrack_bias_resolution_fig_alllengths}
\end{figure}

The {\sc MCTracks} sample is composed of about 53,000 tracks from muons in the momentum range from 0.35 GeV/c to 4.0 GeV/c that have at least one meter of track inside the TPC.  Figure \ref{MCS_true_comparison_exiting_MCTracks_fig} displays the MCS determined momentum versus the true momentum for these muon tracks and figure \ref{MCS_true_exiting_resolution_MCTracks_slices_fig} shows, for six representative bins of true momentum, distributions of $(p_{\text{MCS}}^{-1} - p_{\text{true}}^{-1})/(p_{\text{true}}^{-1})$  along with Gaussian fits to each distribution. These figures indicate that the measured MCS momentum is highly correlated with the true momentum with difference distributions that are well described by a Gaussian distribution with small tails.  The fractional bias and resolution extracted from the Gaussian fits as a function of true momentum are shown in figure \ref{exitingMCTrack_and_RecoTrack_bias_resolution_fig_alllengths}.  The bias is at the few percent level with some slight increase at the higher momenta and the resolution is almost independent of momenta at about the 11\% level.  Since, for this sample, there is no detector measurement uncertainty for the track segments, the results at high momentum are too accurate and will degrade when the expected detector angular measurement uncertainty of 3 mrad is included.  \\

\begin{figure}[ht!]
\centering
	\includegraphics[width=0.9\textwidth]{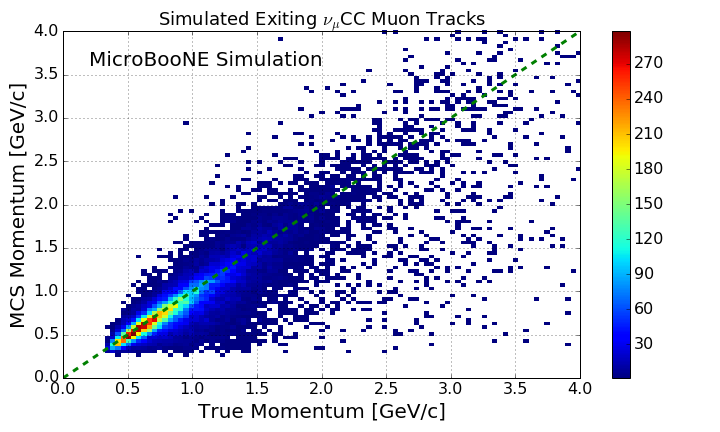} \\
\caption{MCS-computed momentum versus true momentum for the sample of fully-simulated exiting muons from BNB $\nu_\mu$CC interactions in MicroBooNE with at least one meter of track contained within the TPC. The color (z) scale indicates number of tracks.}\label{MCS_true_comparison_exiting_fig}
\end{figure}

The second sample uses a set of muon tracks from fully-simulated BNB $\nu_\mu$CC interactions within the MicroBooNE detector that are automatically reconstructed by the same Pandora algorithm as described in section \ref{track_segmentation_and_scattering_angle_computation_section}.  Again, all tracks are required to have a length of at least $1$~m within the TPC. This simulation does not include space-charge effects. Approximately half of muons from $\nu_\mu$CC interactions within the specified fiducial volume exit the TPC, and about two thirds of those muons have at least one meter of track contained inside of the TPC. The relationship between the reconstructed MCS momentum and the true momentum at the beginning of the track  for this sample of 28,000 exiting muon tracks is shown in figure \ref{MCS_true_comparison_exiting_fig}.  This figure shows a clear core correlation between the MCS and the true momentum but also shows a number of points where the MCS momentum is low with respect to the true momentum.  These low MCS measurements come about because of tracks that are mis-reconstructed.
These mis-reconstructions distort the multiple-scattering angle distribution for the track and push the MCS momentum to lower values especially for high momentum tracks that should have small scattering angles.\\

\begin{figure}
\centering
\includegraphics[width=0.45\textwidth]
	{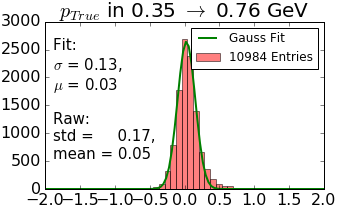}
\includegraphics[width=0.45\textwidth]
	{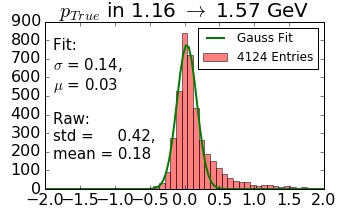}
\includegraphics[width=0.45\textwidth]
	{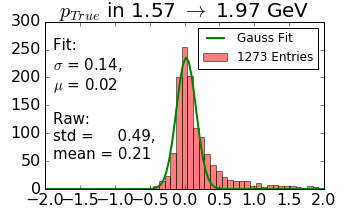}
\includegraphics[width=0.45\textwidth]
	{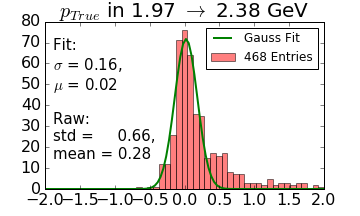}
\includegraphics[width=0.45\textwidth]
	{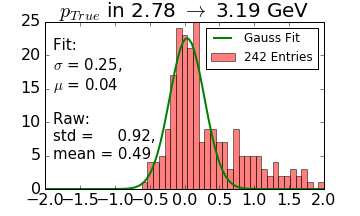}
\includegraphics[width=0.45\textwidth]
	{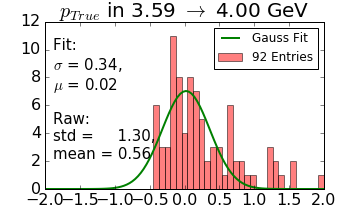}
\caption{Fractional inverse momentum difference for six representative bins of true momentum for the sample of fully-simulated exiting muon tracks. The y-axis is number of tracks, and the x-axis is $(p_{\text{MCS}}^{-1} - p_{\text{true}}^{-1})/(p_{\text{true}}^{-1})$.}
\label{MCS_true_exiting_resolution_MCBNBRecoTrackExiting_slices_fig}
\end{figure}

The fractional momentum distribution, $(p_{\text{MCS}}^{-1} - p_{\text{true}}^{-1})/(p_{\text{true}}^{-1})$,  is shown for six representative bins of true momentum in figure \ref{MCS_true_exiting_resolution_MCBNBRecoTrackExiting_slices_fig}, along with the Gaussian fit to each distribution. These plots quantify the above behavior and show low-momentum, non-Gaussian tails where the MCS momentum is underestimated due to poor track reconstruction for exiting tracks, which is expected to improve with time due to improved algorithms.  To estimate the MCS measurement bias and resolution, one can approximately capture the core distribution of these fractional momentum distributions by fitting a Gaussian to the central region of the distribution.  The mean and width of the Gaussian fits are then used to compute the fractional bias and resolution (respectively).  The mean and standard deviation of the raw data without fitting are also reported in the figure and quantify the large discrepancy introduced by the low MCS measurement tails.\\

Using the Gaussian fit values, the fractional bias and resolution for the reconstructed tracks as a function of true momentum are also shown in figure \ref{exitingMCTrack_and_RecoTrack_bias_resolution_fig_alllengths}.  Reducing the effects of mis-reconstruction by using the Gaussian fit results, the bias is below 4\%  and the resolution is $\sim$14\% for muon momentum below 2 GeV/c , where the low momentum tails are small.  
 Above 2 GeV/c, the low momentum tails become significant but the Gaussian fit results provide an approximation of the mean and standard deviation.  The resolution also worsens for muon momenta above 2 GeV/c because the multiple scattering angles begin to be comparable with the detector segment resolution of 3 mrad. The resolution for exiting tracks improves for longer lengths within the TPC, and, for example,  reaches a 10\% resolution value for muons with $p<2$~$\text{GeV/c}$ with more than 3.5 meters within the TPC. The mean length of track within the TPC for  the muons in this analysis is 212 cm.\\

The consistency of the resolution points shown in figure~\ref{exitingMCTrack_and_RecoTrack_bias_resolution_fig_alllengths} below 2 GeV/c for the {\sc MCTracks} sample versus the fully-simulated track sample validates the use of the Gaussian fit results to approximate the bias and resolution for the fully-simulated sample.  The absence of the low momentum tails at high momentum in the MCTracks distributions shown in figure~\ref{MCS_true_exiting_resolution_MCTracks_slices_fig} as compared to the fully-simulated distributions in figure~\ref{MCS_true_exiting_resolution_MCBNBRecoTrackExiting_slices_fig} demonstrates that these tails are coming from mis-reconstructed tracks 
in the fully-simulated sample.\\

\section{Conclusions}
We have described a multiple Coulomb scattering maximum likelihood method for estimating the momentum of a three dimensional reconstructed track in a LArTPC and have provided motivation for development of such a technique. Using simulation, we have shown that the standard Highland formula should be re-tuned specifically for scattering in liquid argon.  From simulation studies, this tuning improves the fractional bias to below 2\% and reduces the MCS momentum resolution by 30 to 40\% of the untuned value.
After validating range-based momentum-determination techniques with MicroBooNE simulation, we have demonstrated the accuracy and precision of the MCS-based momentum reconstruction in MicroBooNE data by comparing its performance to the range-based method. For 398 fully-contained $\nu_\mu$CC-induced muons from MicroBooNE BNB data, the MCS method exhibits a fractional bias below 3\% and a momentum resolution below 10\%, agreeing with simulation predictions. \\

Two separate samples of simulated uncontained muon tracks in MicroBooNE with at least one meter contained in the active volume have been used to estimate the accuracy of the MCS-based momentum reconstruction for exiting tracks.  For the first truth-based sample of exiting tracks (the {\sc MCTracks} sample), the bias is at the few percent level and the resolution is approximately 11\%.  For a second fully simulated sample of exiting tracks, the comparison of the reconstructed to true momentum shows significant low momentum tails due to mis-reconstructed tracks.  Approximate bias and resolution values for this sample have been estimated by using Gaussian fits to the central core of the momentum difference distributions to reduce the effects from the mis-reconstructions.  Below 2 GeV/c, the values from these fits are consistent with the {\sc MCTracks} results and indicate that a MCS-based reconstruction has the potential for making muon momentum measurements at the 15\% level for exiting muons in MicroBooNE.  Above 2 GeV/c, the results give a first estimate of the MCS momentum measurement capabilities of MicroBooNE for high momentum exiting tracks.\\


\acknowledgments
This material is based upon work supported by the following: the U.S. Department of Energy, Office of Science, Offices of High Energy Physics and Nuclear Physics; the U.S. National Science Foundation; the Swiss National Science Foundation; the Science and Technology Facilities Council of the United Kingdom; and The Royal Society (United Kingdom). Additional support for the laser calibration system and cosmic ray tagger was provided by the Albert Einstein Center for Fundamental Physics. Fermilab is operated by Fermi Research Alliance, LLC under Contract No. DE-AC02-07CH11359 with the United States Department of Energy.


\end{document}